\documentclass[prb,twocolumn,showpacs]{revtex4}
\usepackage{amsmath}
\usepackage{dcolumn}
\usepackage[dvips]{graphicx}

\begin{document}

\title{Temperature dependence of uniform static magnetic susceptibility
in a two-dimensional quantum Heisenberg antiferromagnetic model}
\date{\today}

\begin{abstract}
A perturbation spin-wave theory for the quantum Heisenberg
antiferromagnets on a square lattice is proposed to calculate the
uniform static magnetic susceptibility at finite temperatures,
where a divergence in the previous theories due to an artificial
phase transition has been removed. To the zeroth order, the main
features of the uniform static susceptibility are produced: a
linear temperature dependence at low temperatures and a smooth
crossover in the intermediate range and the Curie law at high
temperatures. When the leading corrections from the spin-wave
interactions are included, the resulting spin susceptibility in
the full temperature range is in agreement with the numerical
quantum Monte Carlo simulations and high-temperature series
expansions.
\end{abstract}

\author{Y. H. Su$^{1}$}
\author{M. M. Liang$^{1}$}
\author{G. M. Zhang$^{2}$}
\affiliation{$^{1}$Department of Physics, Yantai University,
Yantai 264005, China} \affiliation{$^{2}$Department of Physics,
Tsinghua University, Beijing 100084, China}

\maketitle

\section{Introduction}

\label{sec1}

The quantum Heisenberg antiferromagnet (QHAFM) is a prototype model to
describe the magnetic properties in the parent compounds of high temperature
superconducting cuprates.\cite{Manousakis} Non-linear sigma model is shown
that its ground state has long-range order and its low energy excitations
are in a renormalized classical state.\cite%
{ChakravartyPRL,ChakravartyPRB,Chubukov} These results are confirmed by the
quantum Monte Carlo methods\cite{Makivic,Makivicb} and are consistent with
experimental data.\cite{ChakravartyPRL} Moreover, the Schwinger boson mean
field theory\cite{AAPRL,AAPRB} and the variational spin-wave theory\cite%
{Takahashi} have also been developed for the two-dimensional QHAFM, leading
to similar conclusions. Previous theoretical studies have concentrated in
the low temperature region. However, the low-energy interesting physics of
the two-dimensional QHAFM can persist up to the temperature region $T\simeq J
$. Moreover, the anomalous normal state in the charged doped QHAFM in the
cuprates\cite{Timusk} relies on a theory for the magnetic fluctuations in
the intermediate temperature range. Therefore, a theory is needed to be well
defined in the full temperature range.

Although the Schwinger boson mean field theory\cite{AAPRL,AAPRB} and
Takahashi's variational spin-wave theory\cite{Takahashi} for a QHAFM in
square lattices can be extended to the finite temperature range, the mean
field ansatz results in a finite temperature phase transition at $T_{c}=0.91J
$, which explicitly violates the Mermin-Wagner theorem\cite{Mermin}. To
overcome this artificial mean filed divergence of the magnetic
susceptibility, we will propose a perturbation spin-wave theory.

In the Takahashi's variational spin-wave theory, the Dyson-Maleev spin-boson
representation is used for the QHAFM on a square lattice with a N\'{e}el
ordered ground state:
\begin{equation}
S_{i}^{-}=a_{i}^{\dag },S_{i}^{+}=(2S-a_{i}^{\dag
}a_{i})a_{i},S_{i}^{z}=S-a_{i}^{\dag }a_{i},  \label{eqn1.2-a}
\end{equation}%
for the spins in the spin-up sublattice, and
\begin{equation}
S_{j}^{-}=b_{j},S_{j}^{+}=b_{j}^{\dag }(2S-b_{j}^{\dag
}b_{j}),S_{j}^{z}=-S+b_{j}^{\dag }b_{j},  \label{eqn1.2-b}
\end{equation}%
for the spin-down sublattice. The Dyson-Maleev or the Holstein-Primakoff
spin-wave theory works well for the ground state. The corresponding results,
such as the spin-wave velocity, the spin stiffness constant, the sublattice
magnetization, and the perpendicular susceptibility,\cite%
{Igarashi,Canali,Zheng} agree well to the quantum Monte Carlo simulations.%
\cite{Sandvik1997}

To develop a finite temperature spin-wave theory, Takahashi introduced a
constraint for the vanishing local magnetization, i.e.,
\begin{equation}
\langle S_{l}^{z}\rangle =0  \label{eqn1.3}
\end{equation}%
at the $l$'th lattice site, which automatically fulfills the Mermin-Wagner
theorem. Takahashi also introduced a mean field order parameter $\langle
a_{i}b_{j}\rangle =\langle a_{i}^{\dag }b_{j}^{\dag }\rangle $ to account
for the spin-wave interaction in mean field level. This mean field ansatz
can describe reliably the low temperature physics, but fails at high
temperatures, where it produces an artificial phase transition.

In this paper, we present a perturbation spin-wave theory with the
constraint Eq. (\ref{eqn1.3}). The spin-wave interactions are considered
within a many-body perturbation method. The mean field divergence in the
previous variational spin-wave theory is removed. To verify our perturbation
spin-wave theory, we calculate the uniform static susceptibility. Its
temperature dependence in the zeroth order can catch most features given by
the quantum Monte Carlo simulations and high temperature series expansion.
It shows a linear temperature dependence at low temperatures, a smooth
crossover in the intermediate range and the Curie law at high temperatures.
By including the first order corrections from the spin wave interactions,
our perturbation spin-wave theory agrees quantitatively well with the
numerical simulations.

This paper is organized as follows. In Sec. \ref{sec2}, we present
the linearized spin wave results up to the zeroth order. In Sec.
\ref{sec3}, a perturbation theory with the first order corrections
from the spin-wave interactions are developed. We also give the
comparison of the static magnetic susceptibility with the results
of quantum Monte Carlo simulations and high-temperature series
expansions. In Sec. \ref{sec4}, we give the discussion and
summary.

\section{Linearized spin-wave theory}

\label{sec2}

In this paper we mainly discuss a QHAFM on a quare lattice with
Hamiltonian defined by
\begin{equation}
H=J\sum_{\langle ij\rangle }\mathbf{S}_{i}\cdot \mathbf{S}_{j},
\label{eqn2.0}
\end{equation}%
where $J>0$ and $\langle ij\rangle $ denotes nearest-neighbor
sites. Since the ground state of this QHAFM Hamiltonian is an
N\'{e}el ordered state, we separate the square lattice with $2N$
sites into sublattice $A$ for the up spins and sublattice $B$ for
the down spins.

The Dyson-Maleev spin representation given by Eq.(\ref{eqn1.2-a}) and (\ref%
{eqn1.2-b}) are defined upon the N\'{e}el ordered state. In terms
of boson operators, the model can be divided into
\begin{equation}
H=H_{l}+H_{I},  \label{eqn2.1}
\end{equation}%
where $H_{l}$ contains the classical energy and the quadratic terms,
\begin{eqnarray}
H_{l} &=&2JS\sum_{\langle ij\rangle }\left( a_{i}^{\dag }a_{i}+b_{j}^{\dag
}b_{j}+a_{i}b_{j}+a_{i}^{\dag }b_{j}^{\dag }\right)   \notag \\
&&-NJzS^{2},  \label{eqn2.2}
\end{eqnarray}%
and $H_{I}$ contains the quartic terms, describing the spin-wave
interactions,
\begin{equation}
H_{I}=-J\sum_{\langle ij\rangle }a_{i}^{\dag }\left( a_{i}+b_{j}^{\dag
}\right) ^{2}b_{j}.  \label{eqn2.3}
\end{equation}

In this section, we neglect the spin-wave interactions and focus
on the Hamiltonian $H_{l}$. We refer this approximate to a
linearized spin-wave theory (LSW). The spin-wave interaction in
the Hamiltonian $H_{I}$ will be studied in Sec. \ref{sec3}.

The constraint Eq. (\ref{eqn1.3}), from the Dyson-Maleev
representation, can be described by introducing a Lagrangian
Hamiltonian as\cite{Takahashi}
\begin{equation}
H_{\lambda }=-\sum_{i\in A}u_{i}(S-a_{i}^{\dag }a_{i})-\sum_{j\in
B}u_{j}(S-b_{j}^{\dag }b_{j}),  \label{eqn2.4}
\end{equation}%
where the Lagrangian multipliers $\mu _{i}$ and $\mu _{j}$ are
assumed to be site independent, i.e. $\mu _{i}=\mu _{j}=\mu $.
Physically the Lagrangian multiplier $\mu $ plays a role of an
effective chemical potential for the spin waves and leads to a
finite gap, thus an effective mass, for these bosonic excitations
at finite temperatures. For convenience in our following
discussion, we define $ \mu \equiv JzS(\lambda -1)$ with the
coordinate number $z=4$.

By using the Fourier transformations and performing the Bogoliubov
transformations, the Hamiltonian $H_{0}=H_{l}+H_{\lambda }$ can be expressed
as
\begin{eqnarray}
H_{0} &=&\sum_{\mathbf{k}}\varepsilon _{\mathbf{k}}\left( \alpha _{\mathbf{k}%
}^{\dag }\alpha _{\mathbf{k}}+\beta _{\mathbf{k}}^{\dag }\beta _{\mathbf{k}%
}\right) +\sum_{\mathbf{k}}\varepsilon _{\mathbf{k}}  \notag \\
&&-NJz\lambda S(1+2S)+NJzS^{2},  \label{eqn2.5}
\end{eqnarray}%
where $\varepsilon _{\mathbf{k}}=JzS\sqrt{\lambda ^{2}-\gamma _{\mathbf{k}%
}^{2}}$ and $\gamma _{\mathbf{k}}=\cos \frac{1}{2}k_{x}\cos \frac{1}{2}k_{y}$%
. Here $\left( \alpha _{\mathbf{k}}^{\dag },\alpha
_{\mathbf{k}}\right) $ and $\left( \beta _{\mathbf{k}}^{\dag
},\beta _{\mathbf{k}}\right) $ are operators describing two
branches of Bogoliubov excitations. The free energy per site is
given by
\begin{eqnarray}
f &=&\frac{T}{N}\sum_{\mathbf{k}}\ln \left( 2\sinh \left( \frac{\varepsilon
_{\mathbf{k}}}{2T}\right) \right) -\frac{Jz}{2}\lambda S(1+2S)  \notag \\
&&+\frac{JzS^{2}}{2},  \label{eqn2.6}
\end{eqnarray}%
Minimizing the free energy with respect to $\lambda $ leads to a
self-consistent equation
\begin{equation}
S+\frac{1}{2}=\frac{1}{2N}\sum_{\mathbf{k}}\frac{\lambda }{\sqrt{\lambda
^{2}-\gamma _{\mathbf{k}}^{2}}}\coth \left( \frac{\varepsilon _{\mathbf{k}}}{%
2T}\right) ,  \label{eqn2.7}
\end{equation}%
as a consequence of the constraint Eq. (\ref{eqn1.3}).

The uniform static susceptibility per site can be calculated from the static
spin-spin correlation function,
\begin{equation}
\chi (T)=\frac{1}{6NT}\sum_{lr}\langle \mathbf{S}_{l}\cdot \mathbf{S}%
_{l+r}\rangle _{l}.  \label{eqn2.8}
\end{equation}%
Here the thermal average is defined in the zeroth order linearized
spin-wave approximation as $\langle \hat{O}\rangle _{l}\equiv
\frac{Tr\left( e^{-H_{0}/T}\hat{O}\right) }{Tr\left(
e^{-H_{0}/T}\right) }$. It can be easily shown that the transverse
spin-spin correlation function $\chi _{\perp
}=\frac{1}{12NT}\sum_{lr}\langle
S_{l}^{+}S_{l+r}^{-}+S_{l}^{-}S_{l+r}^{+}\rangle =0$. Therefore,
only the
longitudinal spin-spin correlation function $\chi _{z}=\frac{1}{6NT}%
\sum_{lr}\langle S_{l}^{z}S_{l+r}^{z}\rangle $ has a finite value. The
vanishing transverse spin-spin correlation function can also be shown in the
Takahashi's variational spin-wave theory.\cite{Takahashi} Physically it
stems from the local constraint given in Eq. (\ref{eqn1.3}).

The longitudinal spin-spin correlation function can be calculated
by introducing a weak external magnetic field $B\mathbf{e}_{z}$
along $z$-axis which introduces an additional term $H_{B}=-B\left(
\sum_{i\in A}S_{i}^{z}+\sum_{j\in B}S_{j}^{z}\right) $ to the
model Hamiltonian. The longitudinal static susceptibility thus
reads $\chi _{z}=-\frac{\partial ^{2}f(B)}{\partial B^{2}}$, where
the free energy $f(B)$ includes the contribution from the external
field $H_{B}$. Then the uniform static susceptibility within the
linearized spin-wave theory has the form:
\begin{equation}
\chi _{l}(T)=\frac{1}{12NT}\sum_{\mathbf{k}}\sinh ^{-2}\left( \frac{%
\varepsilon _{\mathbf{k}}}{2T}\right) .  \label{eqn2.9}
\end{equation}%
A similar expression has also been obtained by Takahashi in his
variational spin-wave theory.\cite{Takahashi}

\subsubsection{Antiferromagnetic ordered state at zero temperature}

\label{sec2a}

The ground state of the QHAFM on a square lattice has a long-range order\cite%
{Takahashi,Hirsch} due to the presence of Bose-Einstein condensation (BEC)
at the momentum $\mathbf{k}=(0,0)$. Since $\lambda =1+O\left( N^{-1}\right) $
at zero temperature, the spin-wave spectrum is given by
\begin{equation}
\varepsilon _{\mathbf{k}}=JzS\sqrt{1-\gamma _{\mathbf{k}}^{2}}.
\label{eqn2a.1}
\end{equation}%
Comparing to the Takahashi's spin wave spectrum\cite{Takahashi} $\varepsilon
_{MT,\mathbf{k}}=Jzm_{1}\sqrt{1-\gamma _{\mathbf{k}}^{2}}$ with $%
m_{1}=S+0.078974$, we find that it can be obtained by rescaling our
spin-wave result as
\begin{equation}
\varepsilon _{MT,\mathbf{k}}=a\varepsilon _{\mathbf{k}},a=\frac{m_{1}}{S}.
\label{eqn2a.2}
\end{equation}%
The additional factor $a$ is the main difference at low
temperatures between our linearized spin-wave theory and
Takahashi's variational theory .

The spin-spin correlation function in the long distant limit can be derived
as
\begin{equation}
\langle \mathbf{S}_{l}\cdot \mathbf{S}_{l+\mathbf{r}}\rangle =c(\mathbf{r}%
)\left( m_{0}+\frac{1}{2\pi r}\right) ^{2},  \label{eqn2a.3}
\end{equation}%
where $c(\mathbf{r})=1$ if $l,l+\mathbf{r}$ are in the same sublattice, and $%
-1$ if $l,l+\mathbf{r}$ belong to different sublattices.
$m_{0}=S-0.19660$ is the magnetization at zero temperature, which
is the same as that in the Schwinger boson mean field
theory\cite{Hirsch}. Both the spin-spin correlation function and
the local spontaneous magnetization $m_{0}$ in our linearized
spin-wave theory agree with those of the variational spin-wave
theory.\cite{Takahashi} These results imply that, to the order of $O(S^{-1})$%
, the spin-wave interactions have no contributions to the spontaneous
magnetization and static spin-spin correlation at $T=0$K, consistent to the
previous results.\cite{Castilla,Igarashi,Zheng}

\subsubsection{Paramagnetic state at finite temperatures}

\label{sec2b}

\begin{figure}[tbp]
\includegraphics
[angle=0,width=0.95\columnwidth,clip]{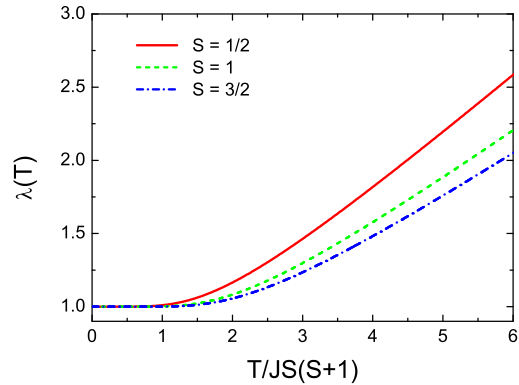}
\caption{The Lagrangian multiplier $\protect\lambda (T)$ in the
linearized spin-wave theory for the QHAFM on a square lattice with
different spin magnitude $S=1/2,1$ and $3/2$. } \label{fig2.1}
\end{figure}

\begin{figure}[tbp]
\includegraphics
[angle=0,width=0.95\columnwidth,clip]{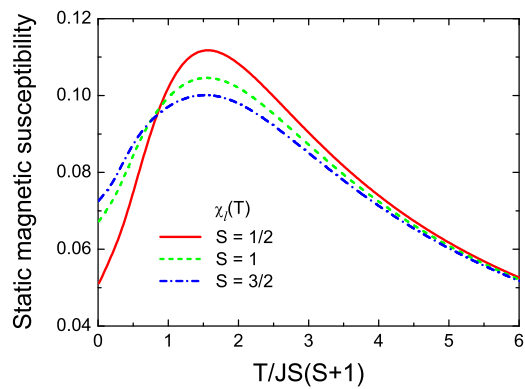}
\caption{The uniform static magnetic susceptibility $\protect\chi
_{l}(T)$ for $S=1/2,1$ and $3/2$ in our linearized spin wave
theory  of the QHAFM on a square lattice. } \label{fig2.2}
\end{figure}

At finite temperatures, according to the Mermin-Wagner theorem, the
long-range order is destroyed in the QHAFM model by quantum fluctuations.%
\cite{Mermin} Fig. \ref{fig2.1} shows the temperature dependence of the
Lagrangian multiplier $\lambda (T)$ with different quantum spins $S=1/2,1,3/2
$. At low temperatures $T\ll JzS$, we follow the Takahashi's method and show
analytically that $\lambda $ approaches exponentially to its BEC value $1$
as
\begin{equation}
\lambda =1+\frac{1}{2}\left( \frac{T}{JzS}\right) ^{2}\exp \left( -\frac{\pi
JzSm_{0}}{T}\right) .  \label{eqn2b.1}
\end{equation}%
The uniform static spin susceptibility in this low temperature region can
also be derived analytically, leading to a linear-temperature dependence,
\begin{equation}
\chi _{l}(T)=\frac{m_{0}}{3JzS}+\frac{2T}{3\pi \left( JzS\right) ^{2}}.
\label{eqn2b.2}
\end{equation}%
The first term gives a finite spin susceptibility at zero
temperature. The linear term results from the linear density of
states of the spin-wave excitations in the low energy limit. The
low-energy spin-wave excitations at low temperatures have an
approximate dispersion
\begin{equation}
\varepsilon _{\mathbf{k}}=JzS\sqrt{\Delta ^{2}+\frac{1}{4}\mathbf{k}^{2}},
\label{eqn2b.3}
\end{equation}%
where a dimensionless bosonic gap is defined by $\Delta \equiv \sqrt{\lambda
^{2}-1}$. At low temperatures, $\lambda $ decays exponentially to $1$ as Eq.(%
\ref{eqn2b.1}), therefore the energy spectrum Eq.(\ref{eqn2b.3}) is
approximately as $\varepsilon _{\mathbf{k}}=\frac{JzS}{2}k$. The density of
states $\rho (\omega )$ of the spin-wave excitations is $\rho (\omega )=%
\frac{1}{(2\pi )^{2}}\int d^{2}\mathbf{k}\delta \left( \omega -\varepsilon _{%
\mathbf{k}}\right) =\frac{2\omega }{\pi \left( JzS\right) ^{2}}$.
It is this linear density of states that leads to the linear-temperature
dependence of the static magnetic susceptibility at low temperatures. Note
that by scaling the linear temperature susceptibility $\chi _{l}(T)$ by the
factor $a$, we can obtain the low-temperature susceptibility given by
Takahashi,\cite{Takahashi} $\chi _{MT}(T)=\frac{1}{a}\chi _{l}\left( \frac{T%
}{a}\right) $.

At low temperatures, the spin-spin correlation function follows
\begin{equation}
\langle \mathbf{S}_{i}\cdot \mathbf{S}_{i+\mathbf{r}}\rangle =c(\mathbf{r})%
\frac{4}{\pi }\left( \frac{T}{JzS}\right) ^{2}\frac{\xi }{r}e^{-r/\xi },
\label{eqn2b.4}
\end{equation}%
where $c(\mathbf{r})$ is defined in Eq.(\ref{eqn2a.3}) and $\xi $ is the
correlation length defined by $\xi \equiv \frac{1}{4\sqrt{\lambda ^{2}-1}}$.
From Eq.(\ref{eqn2b.1}), it can be easily shown that
\begin{equation}
\xi (T)=\frac{JzS}{4T}\exp \left( \frac{\pi JzSm_{0}}{2T}\right) .
\label{eqn2b.5}
\end{equation}%
Takahashi's result for the spin-spin correlation length $\xi _{MT}(T)$ can
be obtained by rescaling with the factor $a$ defined in Eq.(\ref{eqn2a.2}), $%
\xi _{MT}(T)=\xi \left( \frac{T}{a}\right) $. This exponential temperature
dependence of the spin correlation length agrees with that obtained from
one-loop renormalization group approach of the non-linear sigma model.\cite%
{Kopietz,Manousakis} Compared to the experimental data and the numerical
simulations, the two-loop approximation is necessary to convert the $1/T$
prefactor of the exponential into a constant.\cite%
{ChakravartyPRL,ChakravartyPRB}

When increasing the temperature, the Lagrangian multiplier $\lambda (T)$ in
Fig. \ref{fig2.1} evolves from the low-temperature exponential behavior to
the linear temperature dependence at high temperatures. At the same time,
the uniform susceptibility as shown in Fig. \ref{fig2.2} increases firstly
to a broad peak at the intermediate temperature $T_{c}\propto JS(S+1)$, and
then decreases at high temperatures.

At high temperatures $T\gg JzS$, the dispersion of the energy spectrum
becomes relatively weak and $\lambda \gg 1$. The self-consistent equation
Eq.(\ref{eqn2.7}) gives rise to
\begin{equation}
\lambda (T)=\frac{T}{JzS}\ln \left( 1+\frac{1}{S}\right) .  \label{eqn2b.6}
\end{equation}%
In the high-temperature region, the uniform static magnetic susceptibility
obeys the Curie law,
\begin{equation}
\chi _{l}(T)=\frac{(S+1)S}{3T}.  \label{eqn2b.7}
\end{equation}%
Such a high-temperature Curie-law behavior in the linearized
spin-wave theory is exactly the same as that in the Takahashi's
theory.\cite{Takahashi} Why our linearized spin-wave theory can
give rise to a Curie-law behavior, in contrast to the general
spin-wave theory is a very interesting question. The basic reason
is that there is an effective chemical potential introduced for
our spin waves by the local constraint Eq.(\ref{eqn1.3}). This
gives rise to a temperature dependent finite gap for the spin
waves which behave as bosons with effective mass at finite
temperatures.

It should be noted that in our numerical solution to the self-consistent
equation Eq.(\ref{eqn2.7}) and in calculating the uniform magnetic
susceptibility Eq.(\ref{eqn2.9}), there is a trick to include the
contribution of the low-energy spin-wave excitations at low temperatures.
Although no exact BEC occurs at finite temperatures, in the actual numerical
calculation Eq.(\ref{eqn2b.1}) shows us that the exponential decay of the
bosonic gap $JzS\Delta $ at $\mathbf{k}=(0,0)$ will lead to a difficulty in
numerically calculating the momentum integral from the low-energy region $%
\varepsilon _{\mathbf{k}}\ll T$. We call this difficulty as a BEC-like
pseudo-singularity.

A trick to deal with the BEC-like pseudo-singularity is to separate the
momentum space into two regions: Region (i) $\mathbf{k}\in \mathbf{k}%
_{BEC}+\delta _{\mathbf{k}}$ and Region (ii) is the rest part in the first
Brillouin Zone. Here the momentum $\mathbf{k}\in \delta _{\mathbf{k}}$
satisfies $\left| \mathbf{k}\right| <k_{m}$. $k_{m}$ is an irrelevant
parameter and we choose $k_{m}=0.01$ in our calculations. The momentum
integral in Region (i) can be firstly calculated with an approximate
spin-wave energy dispersion Eq.(\ref{eqn2b.3}).

\begin{figure}[tbp]
\includegraphics
[angle=0,width=0.95\columnwidth,clip]{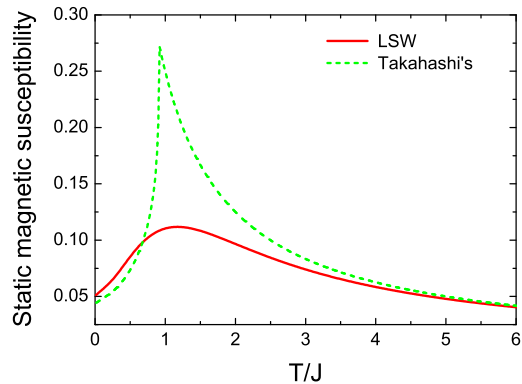}
\caption{Comparison of the uniform static magnetic susceptibility with the $%
S=1/2$ linearized spin-wave theory (LSW) and the Takahashi's variational theory%
\protect\cite{Takahashi} for the QHAFM on a square lattice. }
\label{fig2.3}
\end{figure}

For comparison, the uniform static magnetic susceptibility from our
linearized spin-wave theory and the Takahashi's variational results are
shown in Fig. \ref{fig2.3}. The sharp peak value at $T_{c}=0.91J$ in the
Takahashi's theory comes from an artificial phase transition due to
vanishing of the mean-field ansatz. At low temperatures, the static magnetic
susceptibility from our linearized spin-wave theory is slightly larger than
the Takahashi's result. At high temperatures, the Curie law occurs exactly
in both theories. Clearly, the artificial phase transition has been removed
in the linearized spin-wave theory.

One special feature in the uniform magnetic susceptibility shown in Fig. \ref%
{fig2.2} is a kink structure around the temperature $T\approx 0.5JS$, where
the slop of the susceptibility changes from a smaller value into a larger
one. This special feature is also observed in the quantum Monte Carlo data. %
\cite{Kim1998} It was argued to originate from a crossover from a
renormalized classical regime into a quantum critical regime. However, the
Monte Carlo simulation shows that a similar abrupt change also appears in a
classical Heisenberg model.\cite{Kim1998} Therefore, whether there is a
crossover is still controversial. However, this kink structure implies
obviously different scenarios occurring in these two temperature regions.

\section{First order corrections from the spin-wave interactions}

\label{sec3}

In this section, we will go beyond the linearized approximation by including
the first order corrections from the spin-wave interactions of the QHAFM
model. The method is based on the many-body perturbation theory.

\subsection{Formulation for first order corrections}

After the Fourier and Bogoliubov transformations, the total Hamiltonian $%
H=H_{l}+H_{\lambda }+H_{I}$ can be approximated by\cite{Oguchi}
\begin{eqnarray}
\mathcal{H} &=&\mathcal{H}_{0}+\mathcal{H}_{I}+E_{c},  \notag \\
\mathcal{H}_{0} &=&\sum_{\mathbf{k}}A_{\mathbf{k}}\left( n_{\mathbf{k}}+n_{%
\mathbf{k}}^{\prime }\right) ,  \notag \\
\mathcal{H}_{I} &=&\sum_{\mathbf{k}_{1},\mathbf{k}_{2}}\left[ B^{\left(
1\right) }\left( n_{\mathbf{k}_{1}}n_{\mathbf{k}_{2}}+n_{\mathbf{k}%
_{1}}^{\prime }n_{\mathbf{k}_{2}}^{\prime }\right) +B^{\left( 2\right) }n_{%
\mathbf{k}_{1}}n_{\mathbf{k}_{2}}^{\prime }\right] ,  \notag \\
E_{c} &=&\sum_{\mathbf{k}}\varepsilon _{\mathbf{k}}+NJzS\left[ S-\lambda
\left( 1+2S\right) \right] -\frac{NJz}{4}c_{1}^{2},  \label{eqn4.2}
\end{eqnarray}%
where $n_{\mathbf{k}}\equiv \alpha _{\mathbf{k}}^{\dag }\alpha _{\mathbf{k}}$%
, $n_{\mathbf{k}}^{\prime }\equiv \beta _{\mathbf{k}}^{\dag }\beta _{\mathbf{%
k}}$, $c_{1}(\lambda )=1-\frac{1}{N}\sum_{\mathbf{k}}\frac{\lambda -\gamma _{%
\mathbf{k}}^{2}}{\zeta _{\mathbf{k}}}$, $\zeta _{\mathbf{k}}=\sqrt{\lambda
^{2}-\gamma _{\mathbf{k}}^{2}}$, and $A_{\mathbf{k}}=JzS\zeta _{\mathbf{k}%
}\left( 1+c_{1}\frac{\lambda -\gamma _{\mathbf{k}}^{2}}{2S\zeta _{\mathbf{k}%
}^{2}}\right) $ represents the spin-wave energy spectrum with corrections
from the normal ordering of $H_{I}$. At zero temperature, $\lambda
=1+O\left( N^{-1}\right) $ and the spin-wave energy dispersion is given by $%
A_{\mathbf{k}}=JzS\zeta _{\mathbf{k}}\left( 1+\frac{c_{1}\left( 1\right) }{2S%
}\right) $ with $c_{1}\left( 1\right) =0.15794$ for $S=1/2$. The parameters $%
B^{\left( 1\right) }$ and $B^{\left( 2\right) }$ are defined by
\begin{eqnarray}
B^{\left( 1\right) } &=&-\frac{Jz}{4N}\left( \frac{\left( \lambda -\gamma _{%
\mathbf{k}_{1}}^{2}\right) \left( \lambda -\gamma _{\mathbf{k}%
_{2}}^{2}\right) }{\zeta _{\mathbf{k}_{1}}\zeta _{\mathbf{k}_{2}}}-1\right) ,
\notag \\
B^{\left( 2\right) } &=&-\frac{Jz}{2N}\left( \frac{\left( \lambda -\gamma _{%
\mathbf{k}_{1}}^{2}\right) \left( \lambda -\gamma _{\mathbf{k}%
_{2}}^{2}\right) }{\zeta _{\mathbf{k}_{1}}\zeta _{\mathbf{k}_{2}}}+1\right) .
\notag
\end{eqnarray}

In the above approximation, we have ignored the off-diagonal parts in the
Fock space and picked up all the terms of the two spin-wave interactions.
This is enough when we only consider the first order corrections from the
spin-wave interactions for the free energy, since the linked cluster theorem
tells that $f=f_{0}+\langle \mathcal{H}_{I}\rangle _{0}$, where the thermal
average is defined by $\langle \hat{O}\rangle _{0}\equiv \frac{Tr\left( e^{-%
\mathcal{H}_{0}/T}\hat{O}\right) }{Tr\left( e^{-\mathcal{H}_{0}/T}\right) }$
and $f_{0}$ is defined for the free bosons with energy spectrum $A_{\mathbf{k%
}}$.

The free energy per lattice site reads
\begin{equation}
f=\frac{T}{N}\sum_{\mathbf{k}}\ln \left[ \left( 1-e^{-A_{\mathbf{k}%
}/T}\right) \right] -\frac{Jz}{2}c_{2}^{2}+\frac{1}{2N}E_{c},  \label{eqn3.2}
\end{equation}%
where $c_{2}(\lambda )=\frac{1}{N}\sum_{\mathbf{k}}\frac{\lambda -\gamma _{%
\mathbf{k}}^{2}}{\zeta _{\mathbf{k}}}\tilde{n}_{\mathbf{k}}$ and $\tilde{n}_{%
\mathbf{k}}=\left( e^{A_{\mathbf{k}}/T}-1\right) ^{-1}$. Minimizing this
free energy, we can obtain the following self-consistent equation for the
Lagrangian multiplier $\lambda $,
\begin{eqnarray}
S+\frac{1}{2} &=&\frac{1}{2N}\sum_{\mathbf{k}}\coth \left( \frac{A_{\mathbf{k%
}}}{2T}\right) \left( \frac{\lambda }{\zeta _{\mathbf{k}}}+\frac{\tilde{c}%
\left( \lambda -1\right) \gamma _{\mathbf{k}}^{2}}{2S\zeta _{\mathbf{k}}^{3}}%
\right)   \notag \\
&&+\frac{1}{N}c_{2}\sum_{\mathbf{k}}\frac{\left( \lambda -\gamma _{\mathbf{k}%
}^{2}\right) }{4T\sinh ^{2}\left( \frac{A_{\mathbf{k}}}{2T}\right) S\zeta _{%
\mathbf{k}}}\frac{\partial A_{\mathbf{k}}}{\partial \lambda },
\label{eqn3.2}
\end{eqnarray}%
where $\tilde{c}(\lambda )=c_{1}(\lambda )-2c_{2}(\lambda )$ and
\begin{eqnarray}
\frac{\partial A_{\mathbf{k}}}{\partial \lambda } &=&JzS\left( \frac{\lambda
}{\zeta _{\mathbf{k}}}+c_{1}\frac{(\lambda -1)\gamma _{\mathbf{k}}^{2}}{%
2S\zeta _{\mathbf{k}}^{3}}\right)   \notag \\
&&-Jz\frac{\lambda -\gamma _{\mathbf{k}}^{2}}{2N\zeta _{\mathbf{k}}}\sum_{%
\mathbf{k}_{2}}\frac{(\lambda -1)\gamma _{\mathbf{k}_{2}}^{2}}{\zeta _{%
\mathbf{k}_{2}}^{3}}.  \notag
\end{eqnarray}%
It should be noted that the self-consistent equation
Eq.(\ref{eqn3.2}) is physically equivalent to the first order
perturbation expansion for the constraint Eq.(\ref{eqn1.3}),
\begin{equation}
0=S-\langle a_{i}^{\dag }a_{i}\rangle _{0}+\int_{0}^{1/T}d\tau
\langle T_{\tau }\mathcal{H}_{I}(\tau )a_{i}^{\dag
}(0)a_{i}(0)\rangle _{0}. \label{eqn3.3}
\end{equation}

The uniform static susceptibility can also be calculated from the spin-spin
correlation function given in Eq.(\ref{eqn2.8}). When the first order
corrections from the spin-wave interactions are included, the transverse
static susceptibility has a small but finite contribution,
\begin{eqnarray}
\chi _{\perp }\left( T\right)  &=&\frac{2}{T}\chi _{F}^{2}\left( T\right) ,
\label{eqn3.4} \\
\chi _{F}\left( T\right)  &=&S+\frac{1}{2}+\frac{c_{2}}{2S}\sum_{\mathbf{k}}%
\frac{\left( \lambda -1\right) \gamma _{\mathbf{k}}^{2}}{\zeta _{\mathbf{k}%
_{2}}^{3}}  \notag \\
&&-\frac{1}{2N}\sum_{\mathbf{k}}\coth \left( \frac{A_{\mathbf{k}}}{2T}%
\right) \left( \frac{\lambda }{\zeta _{\mathbf{k}}}+\frac{c_{1}\left(
\lambda -1\right) \gamma _{\mathbf{k}}^{2}}{2S\zeta _{\mathbf{k}}^{3}}%
\right) .  \notag
\end{eqnarray}%
In the derivation for $\chi _{\perp }$, we have used the
perturbation constraint equation Eq.(\ref{eqn3.3}) and ignored the
higher order terms of the spin-wave interactions.

The longitudinal static susceptibility can also be calculated from the free
energy by introducing a weak external field Hamiltonian $H_{B}$, yielding
\begin{equation}
\chi _{z}\left( T\right) =\frac{1}{4NT}\sum_{\mathbf{k}}\sinh ^{-2}\left(
\frac{A_{\mathbf{k}}}{2T}\right) +Jz\left[ c_{2}c_{4}-c_{3}^{2}\right] ,
\label{eqn3.5}
\end{equation}%
where the second term comes from the two spin-wave interaction compared to
the linearized spin-wave expression Eq.(\ref{eqn2.9}), and
\begin{eqnarray}
c_{3}\left( \lambda \right)  &=&\frac{1}{N}\sum_{\mathbf{k}}\frac{-\left(
\lambda -\gamma _{\mathbf{k}}^{2}\right) }{4T\zeta _{\mathbf{k}}\sinh
^{2}\left( \frac{A_{\mathbf{k}}}{2T}\right) },  \notag \\
c_{4}\left( \lambda \right)  &=&\frac{1}{N}\sum_{\mathbf{k}}\frac{\left(
\lambda -\gamma _{\mathbf{k}}^{2}\right) \coth \left( \frac{A_{\mathbf{k}}}{%
2T}\right) }{4T^{2}\zeta _{\mathbf{k}}\sinh ^{2}\left( \frac{A_{\mathbf{k}}}{%
2T}\right) }.  \notag
\end{eqnarray}%
The uniform static susceptibility is thus given by
\begin{equation}
\chi \left( T\right) =\frac{1}{3}\left[ \chi _{\perp }\left( T\right) +\chi
_{z}\left( T\right) \right] .  \label{eqn3.6}
\end{equation}

\subsection{Numerical results}

The self-consistent equation Eq.(\ref{eqn3.2}) and the static magnetic
susceptibility Eq.(\ref{eqn3.6}) have been numerically calculated. When the
first order corrections of the spin-wave interactions are included, the
temperature dependence of both the Lagrangian multiplier $\lambda $ and the
uniform static magnetic susceptibility have similar behaviors to the
linearized spin-wave theory.

\begin{figure}[tbp]
\includegraphics
[angle=0,width=0.95\columnwidth,clip]{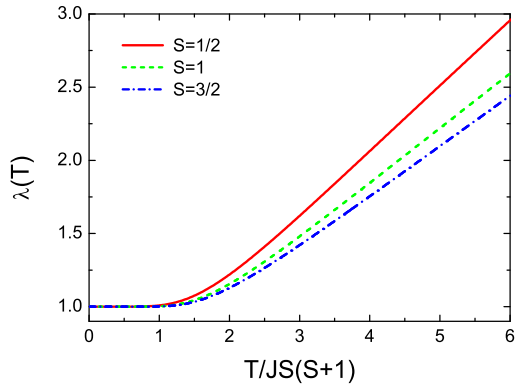}
\caption{The Lagrangian multiplier in our perturbation spin-wave
theory with the first order corrections from the spin-wave
interactions. } \label{fig3.1}
\end{figure}

The Lagrangian multiplier $\lambda $ is shown in Fig. \ref{fig3.1} for
different spin $S=1/2,1$ and $3/2$. At low temperatures, it decays
exponentially to its zero-temperature value $1$, indicating the gapless spin
wave excitations. At high temperatures, $\lambda $ has a linear-temperature
variation $\lambda (T)=\kappa _{0}+\kappa _{1}T$, where $\kappa _{0}$ and $%
\kappa _{1}$ have weak temperature dependence. For $S=1/2$, $\kappa
_{0}=0.334$ and $\kappa _{1}=0.583$ in the temperature region $3J<T<10J$.

\begin{figure}[tbp]
\includegraphics [angle=0,width=0.95\columnwidth,clip]{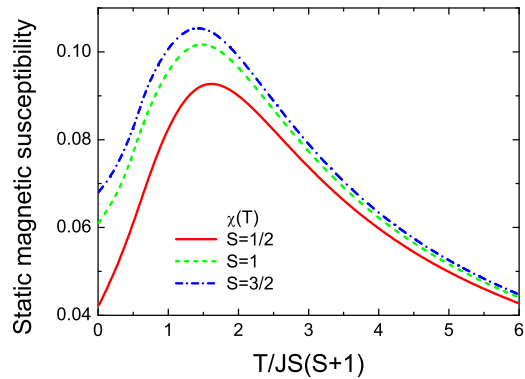}
\caption{The static magnetic susceptibility with different spin
$S=1/2,1$ and $3/2$ in our perturbation spin-wave theory with the
first order corrections from the spin-wave interactions. }
\label{fig3.2}
\end{figure}

The static magnetic susceptibility is displayed in Fig. \ref{fig3.2}. It has
three main features: (i) the linear-temperature behavior at low
temperatures, (ii) a broad peak at an intermediate characteristic
temperature $T_{c}\propto JS(S+1)$, (iii) the Curie-Weiss law at high
temperatures. For $S=1/2$, a low-temperature linear behavior can be fitted
by $\chi (T)=\chi _{c}+bT$, with $\chi _{c}=0.0418$ and $b=0.0471$. Compared
to the Takahashi's theory where a similar linear behavior with $\chi
_{c}=0.0437$ and $b=0.0396$,\cite{Takahashi} our zero temperature
susceptibility $\chi _{c}$ is slightly smaller than that of Takahashi's.
Also the slope $b$ is also slightly larger than that of the Takahashi's.

\begin{figure}[tbp]
\includegraphics [angle=0,width=0.95\columnwidth,clip]{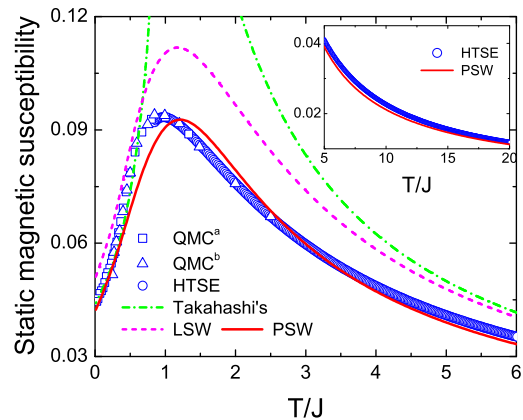}
\caption{Comparison of the uniform static susceptibilities from
the linearized spin-wave theory (LSW) and the perturbation
spin-wave theory (PSW) with first order correction for the QHAFM
on a square lattice. We also show the result from the Takahashi's
modified spin-wave theory and numerical data from Quantum Monte
Carlo and high-temperature series expansions (HTSE). The
inset shows the comparison of PSW with HTSE in high temperature region. QMC$%
^{a}$ comes from Ref. \onlinecite{Kim1998} and QMC$^{b}$ comes
from Ref. \onlinecite{Makivic}. HTSE comes from Eq.($34$) of Ref.
\onlinecite{Takahashi}. For comparison, QMC$^{b}$ have been
multiplied by a factor $\frac{1}{3}$. } \label{fig3.3}
\end{figure}

At high temperatures, the spin-wave interactions modify the exact Curie law
of the static magnetic susceptibility Eq.(\ref{eqn2b.7}) into a Curie-Weiss
behavior. As shown in Fig. \ref{fig3.3}, it evolves into the behavior
obtained from the numerical high-temperature series expansions. Physical
origin of the high-temperature Curie-Weiss law also comes from the
linear-temperature dependence of Lagrangian multiplier $\lambda $ in the
energy spectrum of the spin waves.

It should be noted that the kink structure in the static magnetic
susceptibility in the linearized spin-wave theory shown in Fig. \ref{fig2.2}
also occurs in the perturbation theory. This implies that the kink feature
is a characteristic property in the QHAFM model.

In Fig. \ref{fig3.3}, we compare the static magnetic susceptibility of our
spin-wave theory to the quantum Monte Carlo data\cite{Kim1998,Makivic} and
the high-temperature series expansions. It shows that our linearized
spin-wave theory can reproduce all the main features in the numerical
simulations: the low temperature linear behavior, the smooth crossover in
the intermediate temperature region and the Curie-Weiss law behavior at high
temperature. This good agreement encourages us to study the frustrated $%
J_{1}-J_{2}$ model proposed to describe the magnetic fluctuation in the
newly discovered iron-pnictide superconductors.\cite{Zhang} The first order
correction from the spin-wave interactions makes the uniform static
susceptibility both at low and high temperatures agrees quantitatively with
the numerical quantum Monte Carlo and high temperature series expansions.
This good fitting in Fig. \ref{fig3.3} indicates that the physics of QHAFM
model, at least to the static magnetic response, has been captured by our
perturbation spin-wave theory in the full temperature region.

\section{Discussions and Conclusions}

\label{sec4}

The good agreement of our perturbation spin-wave theory with the
numerical simulations shows the validity of our theory at finite
temperatures. It seems surprising because the spin-wave theory
developed in the pioneering
papers\cite{Bloch,Holstein,Anderson,Kubo} is generally thought to
work well only at low temperatures in the ordered phase. The main
difference of our finite temperature spin-wave theory is the
requirement of the vanishing local magnetization. It is this local
constraint that restricts the number of the spin wave excitations
which thus behaves as bosons with an effective mass at finite
temperatures.

The extension of the spin-wave theory to the low-dimensional spin
systems have been shown to work well. In one-dimensional quantum
magnets, no magnetic order can survive at zero temperature due to
the strong quantum fluctuations. Although there is no long range
order in the ground state, the free energy and the static magnetic
susceptibility of the one-dimensional ferromagnet from a modified
spin-wave theory show same behaviors to Bethe-Ansatz
solutions.\cite{TakahashiFM} The Haldane gap for the
one-dimensional quantum antiferromagnet with $S=1$ from a
spin-wave theory is $\Delta _{H}=0.3914J$ (Ref.
\onlinecite{WangSu}) which well agrees with the numerical
density-matrix renormalization group ($0.4105J$ from Ref.
\onlinecite{White}). The spin-wave theory for the one-dimensional
quantum antiferromagnet with integral spin can reproduce the
static magnetic susceptibility which agrees well with the quantum
Monte Carlo simulations.\cite{Yamamoto} These previous results,
together with our study, indicate that the formulation of the
spin-wave theory is a valid formalism to apply to the
low-dimensional quantum magnets.

In summary, we have studied the QHAFM on a square lattice by using
a perturbation spin-wave theory. The Lagrangian constraint for the
absence of the local magnetization plays an important role and
leads to an effective mass for the spin wave excitations at finite
temperatures. In linearized spin-wave theory, the calculation of
the uniform static magnetic susceptibility can reproduce almost
all features obtained from the quantum Monte Carlo and the
high-temperature series expansions. It has a linear-temperature
dependence at low temperatures and a broad peak feature at an
intermediate characteristic temperature $T_{c}\propto JS(S+1)$. At
high temperatures, the uniform magnetic susceptibility follows a
Curie-Weiss law. The static magnetic susceptibility from our
perturbation theory with the first corrections from the spin-wave
interactions agrees well with the numerical simulations.

\begin{acknowledgments}
Y. H. gratefully acknowledges valuable discussions with Dr. Fei Ye and Prof.
Tao Li. This work is partially supported by NSFC-China and the National
Program for Basic Research of MOST, China.
\end{acknowledgments}



\end{document}